\documentclass[a4paper,preprintnumbers,floatfix,superscriptaddress,aps,prl,twocolumn,showpacs,notitlepage,nofootinbib,nobibnotes]{revtex4-2}
\usepackage[utf8]{inputenc}
\usepackage[T1]{fontenc}
\usepackage[sc,osf]{mathpazo}
\usepackage{amsmath, amsthm, amssymb, amsfonts, mathbbol, amstext}
\usepackage{graphicx}
\usepackage{dcolumn}
\usepackage{bm}
\usepackage{bbm}
\usepackage[colorlinks=true,citecolor=Green,linkcolor=Purple,urlcolor=Blue]{hyperref} 
\usepackage{mathtools}
\usepackage{color}
\usepackage{multirow}
\usepackage{siunitx}
\usepackage{physics}

\usepackage[xcolor=dvipsnames]{changes}
\newcommand{\id}{\mathbb{1}}
\usepackage{soul}
\usepackage{tcolorbox}
\renewcommand{\tr}{\mathrm{tr}}
\renewcommand{\H}{\mathcal{H}}

\newcommand{\dket}[1]{\vert#1\rangle\hspace{-.8mm}\rangle}
\newcommand{\dketbra}[2]{\vert #1 \rangle \hspace{-.8mm} \rangle \hspace{-.4mm} \langle\hspace{-.8mm}\langle #2 \vert}

\newcommand{\PRLsep}{\noindent\makebox[\linewidth]{\resizebox{0.625\linewidth}{1pt}{$\bullet$}}\bigskip}

\definecolor{teocolbox}{RGB}{244,250,251}

\usepackage[normalem]{ulem}

\begin{document}
\title{Experimental superposition of a quantum evolution with its time reverse}
\author{Teodor Strömberg}\email[Corresponding author: ]{teodor.stroemberg@univie.ac.at}
\author{Peter Schiansky}\affiliation{University of Vienna, Faculty of Physics \& Vienna Doctoral School in Physics,  Boltzmanngasse 5, A-1090 Vienna, Austria}
\author{Marco Túlio Quintino}
\affiliation{Sorbonne Universit\' {e}, CNRS, LIP6, F-75005 Paris, France}
\affiliation{University  of Vienna, Faculty of Physics \& Research Network Quantum Aspects of Space Time (TURIS), Boltzmanngasse 5, 1090 Vienna, Austria}
\affiliation{Institute for Quantum Optics and Quantum Information, Boltzmanngasse 3, 1090 Vienna, Austria}
\author{Michael Antesberger}
\affiliation{University of Vienna, Faculty of Physics \& Vienna Doctoral School in Physics,  Boltzmanngasse 5, A-1090 Vienna, Austria}
\author{Lee Rozema}
\affiliation{University of Vienna, Faculty of Physics \& Research Network Quantum Aspects of Space Time (TURIS), Boltzmanngasse 5, 1090 Vienna, Austria}
\author{Iris Agresti}
\affiliation{University of Vienna, Faculty of Physics \& Research Network Quantum Aspects of Space Time (TURIS), Boltzmanngasse 5, 1090 Vienna, Austria}
\author{\v{C}aslav Brukner}
\affiliation{Institute for Quantum Optics and Quantum Information, Boltzmanngasse 3, 1090 Vienna, Austria}\affiliation{University of Vienna, Faculty of Physics \& Research Network Quantum Aspects of Space Time (TURIS), Boltzmanngasse 5, 1090 Vienna, Austria}
\author{Philip Walther}\email[Corresponding author: ]{philip.walther@univie.ac.at}\affiliation{University of Vienna, Faculty of Physics \& Research Network Quantum Aspects of Space Time (TURIS), Boltzmanngasse 5, 1090 Vienna, Austria}

\begin{abstract}
\textbf{In the macroscopic world, time is intrinsically asymmetric, flowing in a specific direction, from past to future. However, the same is not necessarily true for quantum systems, as some quantum processes produce valid quantum evolutions under time reversal. Supposing that such processes can be probed in both time directions, we can also consider quantum processes probed in a coherent superposition of forwards and backwards time directions. This yields a broader class of quantum processes than the ones considered so far in the literature, including those with indefinite causal order. In this work, we demonstrate for the first time an operation belonging to this new class: the quantum time flip. Using a photonic realisation of this operation, we apply it to a game formulated as a discrimination task between two sets of operators. This game not only serves as a witness of an indefinite time direction, but also allows for a computational advantage over strategies using a fixed time direction, and even those with an indefinite causal order.}
\end{abstract}

\maketitle

\section{Introduction}
In recent years, the framework of quantum theory has been generalised to describe agents interacting through quantum processes with indefinite causal orders~\cite{hardy2005probability,chiribella2013quantum,oreshkov2012quantum}. These processes have been realised experimentally using photonic platforms~\cite{procopio2015experimental,rubino2017experimental, rubino2022experimental, goswami2018indefinite, rubino2021experimental}, thereby witnessing the implementation of causally non-separable series of events. Remarkably, these are not the most general processes allowed by quantum mechanics. Take, for example, the quantum SWITCH~\cite{chiribella2013quantum}: even though the causal order of the constituent events is indefinite, each operation is accessed only in a single time direction. By considering processes where the time direction of the underlying operations is indefinite, one can go beyond the framework of indefinite causality.
Indeed, a quantum superposition of evolutions with opposite thermodynamic arrows of time was first proposed in~\cite{rubino2021quantum}.

Processes with an indefinite time direction can be studied by considering operations that exhibit a time symmetry; these operations admit a change of reference frame that yields a valid quantum evolution in which the time coordinate is inverted.
Unitary channels are an example of such operations, and in particular they admit the following time-reversal symmetries: for every evolution $U$, both the inverse $U \mapsto U^{-1}$ and the transpose $U \mapsto U^T$ are valid time-reversal operations.
The presence of such a symmetry naturally excludes evolutions with an arrow of time, such as the thermodynamic processes studied in Ref.~\cite{rubino2021quantum}
    
Given quantum operations that can in principle be accessed in both time directions, we can consider coherent superpositions of transformations made in the forwards and backwards time-directions. This amounts to a new kind of process, which we will refer to as being \textit{inseparable in its time direction}, an example of which - called the \textit{quantum time flip} - was recently introduced in \mbox{Ref.~ \cite{chiribella2020quantum}}.
    \begin{figure*}[th]
    \centering
        \includegraphics[width=\linewidth]{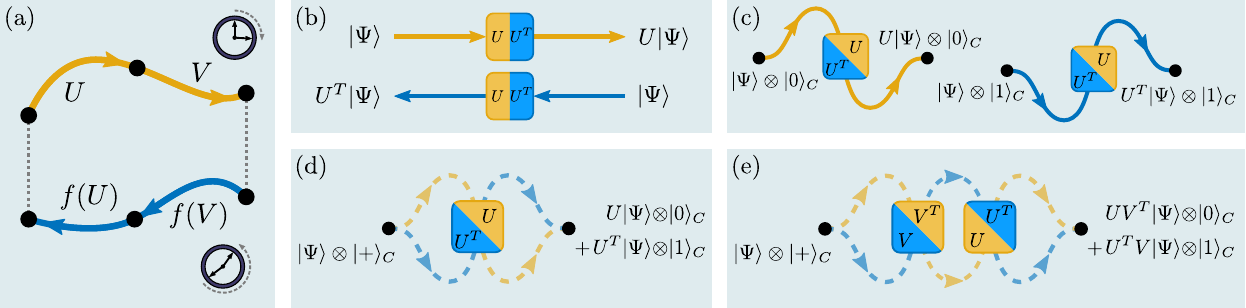}
        \caption{\textbf{Time-reversal and the quantum time flip.} \textbf{(a)} The forwards (top) and backwards (bottom) directions of the same time-evolution are shown in yellow and blue, respectively. The backwards time-evolution is given by some function $f$ of the forwards evolution, and decomposing the total time evolution into steps shows that $f$ must be order reversing. The inverse and transpose are examples of such order reversing functions.
        \textbf{(b)} Quantum gates are often modelled as black boxes with an input and an output. In this work, we consider black boxes that can be accessed in two different directions, producing either the forwards or backwards time-evolution depending on which direction the box is accessed in. Here, the backwards time-evolution is taken to be the transpose.
        \textbf{(c)} A control degree of freedom can be introduced to control in which direction the black box is accessed.
        \textbf{(d)} By putting the control qubit in a coherent superposition of the two states in (c) the box is accessed in a superposition of both directions, and the input state is propagated in a superposition of time directions. This is a realisation of the quantum time flip. \textbf{(e)} The quantum time flip can be applied to more than a single gate. This figure illustrates a scenario where two gates are accessed in a superposition of orders, in which they always have the opposite time directions. As described in the main text, this use of two quantum time flips can yield a computational advantage.
        }
        \label{fig:fig1}
    \end{figure*}
This process cannot be realised within the quantum circuit model. In this work we nevertheless present a photonic implementation of the quantum time flip by exploiting \textit{device dependent symmetries} of our experimental apparatus. A quantum state undergoing a time evolution is encoded in the polarization degree of freedom of a single photon, while a control qubit determining the time direction is encoded in its path degree of freedom. 
We show that polarization operations with waveplates naturally implement different time directions for forwards and backwards propagation directions through the waveplates, given the correct Stokes-parameter convention.
This results in a deterministic time-reversal, in contrast to more general approaches which may involve multiple uses of the input operation in combination with probabilistic or non-exact methods~\cite{ebler16,sardharwalla16,navascues17,quintino19PRL,quintino19PRA,quintino21unitary,trillo19,trillo22,Schiansky22,yoshida22}.
We can furthermore realise the quantum time flip deterministically by passing the photon through the waveplates in a superpositon of the two propagation directions.

We certify the indefinite time direction by demonstrating an information-theoretic advantage of the quantum time flip in the context of a computational game. In this setting, the quantum time flip not only outperforms strategies that utilise operations with a fixed time direction, but even strategies that exploit operations with an indefinite causal order~\cite{procopio2015experimental,taddei2021computational}.

\section{Quantum circuits, unitary transposition, and processes with indefinite time direction}
The standard quantum circuit formalism provides solid grounds for quantum computing and forms the basis for quantum complexity theory~\cite{NilsenChuangBook,watrous_complexity}. However, it also imposes limitations on how we apply quantum theory. In a circuit, operations necessarily respect a definite causal order and the strict notion of input and output.
{The existence of time reversal processes such as unitary transposition is forbidden by the standard circuit formalism when given access to one~\cite{quintino19PRA,quintino21unitary} or even two~\cite{chiribella2020quantum} uses of an unknown unitary. However, for practical and foundational reasons, researchers have been designing and pursuing non-exact and probabilistic schemes aimed towards this goal~\cite{ebler16,sardharwalla16,navascues17,quintino19PRA,quintino19PRL,quintino21unitary,trillo19,trillo22,Schiansky22}. Remarkably, a very recent work shows that in the qubit case, when four uses of the input operation are available, there exists a quantum circuit to invert arbitrary unitary operations \cite{yoshida22}.}
    	
In quantum theory, reversible operations are described by unitary operators. Processes which reverse a composition of such operations may be expressed by a function $f$ satisfying
\begin{equation}
	 f(UV)=f(V)f(U), \quad \forall U,V,
\end{equation}
for all unitary operators $U$ and $V$ (see Fig.~\ref{fig:fig1}). Under natural assumptions, it can be proven that, up to a unitary transformation, there are only two time reversal functions $f$,  unitary transposition $f(U)=U^T$ and unitary inversion $f(U)=U^{-1}$~\cite{chiribella2020quantum}. For two-dimensional systems, unitary transposition and unitary inversion are unitarily equivalent via a Pauli $\sigma_Y$ operation. This follows from the identity, $U^{-1}=\sigma_Y U^T \sigma_Y$ which holds for all operators $U\in\mathcal{SU}(2)$. Hence, for qubits, universal unitary transposition is possible if and only if unitary inversion is possible. This equivalence does not hold for higher-dimensional systems, and in these cases the transpose is singled out as the only time-reversal operator for which the quantum time flip is defined~\cite{chiribella2020quantum}. Together with the fact that in the Choi-matrix formalism the transpose has the interpretation of exchanging the roles of input and output of a channel, this motivates the choice of the transpose as time-reversal operator for qubit systems.

\begin{figure*}
    \includegraphics[width=1\linewidth]{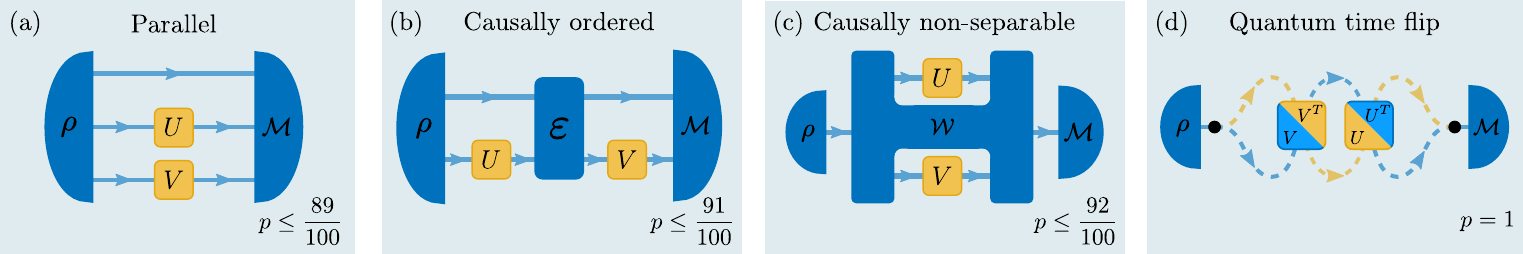}
    \centering
    \caption{\textbf{Classes of game strategies.}
    The figure depicts the different strategies for the game described in the main text and their corresponding maximum winning probabilities $p$. These maximum winning probabilities are obtained through an optimization over all possible choices of the resources shown in dark blue, and hold for pairs of unitaries $(U,V)$ uniformly randomly picked from the sets $\mathcal{M}_+$ and $\mathcal{M}_-$. The state $\rho$, for example, is allowed to contain any number of auxiliary degrees of freedom, and analogous statements hold for the measurement $\mathcal{M}$, channel $\mathcal{E}$ and process $\mathcal{W}$. The three strategies differ in how they are able to access the gates picked by the referee. The strategies (a)-(c) are shown here in the forwards time direction, but are also valid in the backwards time direction in which both gates are transposed. Each subsequent strategy is strictly better than the previous one, and only players who have access to a quantum time flip process can win the game with unity probability. \textbf{(a)} Parallel gate order. \textbf{(b)} Causally ordered gate sequence. \textbf{(c)} Process without a definite causal order. \textbf{(d)} Quantum time flip. }
    \label{fig:figMTQ}
\end{figure*}
When focusing on a particular physical implementation, the general aspects of the standard quantum circuit formalism may limit our view and lead to an apparent mismatch between theory and practice. A known illustrative example is the universal coherent control of unitary operations, where an arbitrary unitary $U$ is applied to the target system conditional on the state of a control qubit: $U\mapsto \id\otimes \ketbra{0}{0}_C+U\otimes \ketbra{1}{1}_C$. While it is not possible to design a quantum circuit to perform universal control, a simple Mach-Zehnder optical interferometer can be used for this task~\cite{soeda13,araujo2014quantum,bisio16}. Indeed, experimental control of black box quantum gates has been demonstrated \cite{zhou2011adding, thompson2018quantum}. Such experimental implementations exploit the knowledge of the position of the physical device performing the gate, circumventing this apparent limitation imposed by the quantum circuit formalism. 

Although time reversal processes such as unitary transposition are not possible within the standard circuit formalism when given access to one~\cite{quintino19PRA,quintino21unitary} or even two~\cite{chiribella2020quantum} uses of an unknown unitary, in this work we implement general qubit unitary transposition, as well as the quantum time flip process, using a particular optical construction.
Similar to the case of universal coherent control, we make use of knowledge about our specific experimental apparatus to realise a black box unitary that may be used in two different directions. As shown in  Fig.~\ref{fig:fig1}.b, this box implements $U$ in the `forwards' direction, while in the `backwards' direction it has the effect of the transposed operation $U^T$.

%

Moreover, in addition to ``simply'' reversing a quantum evolution, we also coherently superpose the forwards and backwards time evolutions, and in so doing perform an optical implementation of a process with an indefinite time direction \cite{chiribella2020quantum}, i.e. one which cannot be described as a convex mixture of processes in which each gate is accessed only in one time direction. The process that we implement optically is the quantum time flip for unitary transposition, a process which acts on unitary operations as
\begin{equation}\label{eq:time_flip}
	U \mapsto U\otimes\ketbra{0}{0}_C + U^T\otimes\ketbra{1}{1}_C.
\end{equation}
We then compose the time flip process of Eq.~\eqref{eq:time_flip} with its flipped version, $V \mapsto V^T\otimes\ketbra{0}{0}_C + V\otimes\ketbra{1}{1}_C$, to obtain a process which acts on a pair of unitary operators as
\begin{equation}
    \label{eq:time_flip2}
	(U,V) \mapsto UV^T\otimes\ketbra{0}{0}_C + U^TV \otimes\ketbra{1}{1}_C.
\end{equation}
In addition to having an indefinite time direction, the process described in Eq.~\eqref{eq:time_flip2} cannot be described by general process matrices with indefinite causality such as the quantum switch~\cite{chiribella2013quantum} or the Oreshkov-Costa-Brukner (OCB) process~\cite{oreshkov2012quantum}. In the next section, we will explain how to witness this property.
    
\section{Game description}
We now describe a discrimination task, first introduced in Ref.~\cite{chiribella2020quantum}, where the quantum time flip process will be used as a resource to increase our performance.
In this game, a referee provides the player with two black box unitaries, $U$ and $V$, belonging to either the set 
$\mathcal{M}_+$ or $\mathcal{M}_-$, which are known to respect the property
\begin{align}
     \mathcal{M}_+:=&\Big\{(U,V): UV^T= + U^TV \Big\}   \label{c1} \\
     \mathcal{M}_-:=&\Big\{(U,V): UV^T= -U^TV \Big\}.  \label{c2}
\end{align}
The player is then challenged to determine which of the two sets the gates were picked from, while only being allowed to access each of the black boxes once.
    
As discussed in the previous section, a player able to perform the quantum time flip may implement the process in Eq.~\eqref{eq:time_flip2}.
Consider as a strategy an initial state of the form $|\psi\rangle_T \otimes |+\rangle_C$, where $|\pm\rangle_C=\frac{|0\rangle_C \pm |1\rangle_C}{\sqrt{2}}$, $|\psi \rangle$ is an arbitrary state, and the subscripts $C$ and $T$ refer to the control and target qubits. Sending this state through the gate in Eq.~\eqref{eq:time_flip2} gives the state
\begin{equation}
    \bigg[\frac{UV^T+U^TV}{2}\bigg]\ket{\psi}_T |+\rangle_C  +  \bigg[\frac{UV^T-U^TV}{2}\bigg]\ket{\psi}_T  |-\rangle_C.
\label{eq:quantum_flip2}
\end{equation}
Since the states $\ket{\pm}$ are orthogonal, a player using this strategy can always correctly determine which set was chosen by the referee.

In contrast, players who do not have access to indefinite time strategies may not be able to ascertain with certainty to which set a given pair of unitaries $(U,V)$ belongs.
In order to make this claim concrete, Ref.~\cite{chiribella2020quantum} considers a particular game where the set $\mathcal{M}_+$ has 13 pairs of unitary operators respecting $UV^T= + U^TV$, and $\mathcal{M}_-$ has 8 pairs of unitary operators respecting $UV^T= - U^TV$; these two sets of unitary operators are presented in \hyperref[fig:box1]{Box~1}. Here, we consider an average case variation of the aforementioned game, which goes as follows: with uniform probability $p=\frac{1}{13+8}$, the referee picks a pair of unitary operators $(U,V)$ from $\mathcal{M}_+$ or $\mathcal{M}_-$ and lets the player make a single use of each. We then consider the optimal success probability of players who have access to different kinds of resources. As indicated by Eq.~\eqref{eq:quantum_flip2}, players who have access to the quantum time flip can always win with unity probability.
The three other classes of strategies, shown in Fig.~\ref{fig:figMTQ}, only have access to a single time direction, forwards or backwards, and convex combinations of these strategies will be called separable in their time direction; a detailed mathematical characterisation of these strategies is presented in the methods. Employing the computer-assisted proof methods of Ref.~\cite{bavaresco21} we obtain upper bounds on the maximal success probabilities for players restricted to particular classes of strategies. The code for this is openly available in our online repository, see Methods for details.
    
The first alternative strategy we consider is one in which the player is restricted to using $U$ and $V$ in parallel, and this results in a maximal success probability that is bounded by ${\frac{88}{100}\leq}p_\text{par}\leq \frac{89}{100}$. Next, we consider players restricted to causally ordered strategies, whose maximal success probability is found to be bounded by $\frac{90}{100}\leq p_\text{causal}\leq \frac{91}{100}$. Finally, players given access to process matrices with indefinite causality (also called indefinite testers~\cite{chiribella18cause_effect}), but with definite time direction, have their maximal success probability bounded by $\frac{91}{100}< p_\text{i.c.}\leq \frac{92}{100}$. Unlike the task in~\cite{abbott2020communication}, in which causally ordered and general non-quantum-circuit-model strategies perform equally well, this game is hence an example of a channel discrimination task with strict hierarchy between four different classes of strategies. Additionally, while the operations selected by the referee 
are treated as being fully characterised in the above analysis, there are no assumptions made about the measurements performed by the player, and these can remain unknown.
This is therefore an example of a semi-device-independent certification of an indefinite time direction~\cite{bavaresco19,cao22}. This stands in contrast to witness based approaches, previously used to certify advantages in channel discrimination tasks~\cite{rubino2017experimental}, in which one needs well characterised measurement devices in order to evaluate the witness operator.

\section{Experiment}
\begin{figure*}[!ht]
    \includegraphics[width=1\linewidth]{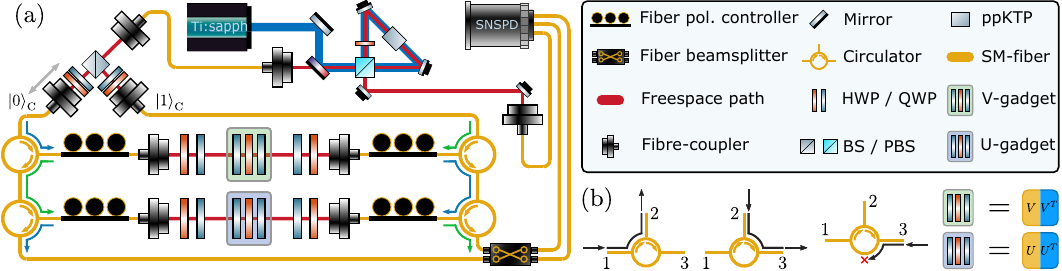}
    \centering
    \caption{\textbf{Experimental apparatus.}
    \textbf{(a)} A type-II spontaneous parametric down-conversion source generates frequency degenerate single-photon pairs at \SI{1546}{\nano\meter} in a ppKTP crystal (top). The signal photon is sent to a heralding detector, while the idler photon is routed to a balanced bulk beam-splitter and coupled into single-mode fiber. Blue (green) arrows indicate the photon path corresponding to the control-qubit state $\ket{0}_C$ ($\ket{1}_C$). A piezo-electric actuator attached to one of the fiber couplers allows for control over the interferometric phase, pairs of HWPs / QWPs are used in combination with fiber polarization controllers for polarization compensation through the fibers. This use of redundant polarization elements both improves and simplifies the compensation procedure~\cite{stromberg2023prescriptive} After the initial beam-splitter two fiber circulators guide the photon through the V-gadget. 
    Propagating through the gadget in the `forwards' direction implements the unitary $V$, while propagating `backwards' has the effect of applying the transposed operation relative to the `forward' direction. One of the two  paths through first the gadget therefore results in the $V^{T}$ being applied instead of $V$. Two additional circulators then route the photon through a gadget implementing $U$ ($U^{T}$) in `forwards' (`backwards') direction. Finally, the signal photon is sent to a fiber beam-splitter, which applies a Hadamard gate on the path degree of freedom, and correlates the two spatial output modes with the sets $\mathcal{M}_+$, $\mathcal{M}_-$. Detection is performed by superconducting nanowire single-photon detectors (SNSPDs) housed in a \SI{1}{\kelvin} cryostat. Additional QWP/HWP pairs are used to compensate fiber-induced polarization rotations. \textbf{(b)} The fiber circulators route the light from port $1\rightarrow 2$, from $2 \rightarrow 3$ and block light entering in port 3. The bidirectional boxes in Fig.~\ref{fig:fig1} are realised using sets of three waveplates. Depending on the propagation direction, they implement either the unitary operation $U$/$V$ or $U^T$/$V^T$.}
    \label{fig:setup}
\end{figure*}

Our photonic implementation of the game described in the previous section makes use of the quantum time-flip strategy from Eq.~\eqref{eq:quantum_flip2} to achieve a success probability exceeding that of any strategy only using the gates in one time direction. To coherently apply the quantum time flip, we employ polarization optics in a partially common-path interferometer, depicted in Fig.~\ref{fig:setup}, with the control and target qubits being encoded in the path and polarization degrees of freedom of a single-photon, respectively. Our experiment makes use of two quantum time flips, sequentially applied to the two unitaries $V$ and $U$. The resulting controlled channel is the one of Eq.~\eqref{eq:quantum_flip2} where the gates $UV^T$ and $U^TV$ act on the target (polarization) qubit and are implemented using two Simon-Mukunda polarization gadgets consisting of three waveplates each~\cite{simon1990minimal}, for which the transpose operation is obtained by reversing the propagation direction.

Such polarization gadgets generally do not realise the transpose operation in the backwards propagation direction, but rather a related operation:
\begin{equation}
    U_{\mathrm{fw}} \mapsto U_{\mathrm{bw}} = PU_{\mathrm{fw}}^T P^{\dagger},
\end{equation}
where $P$ is a matrix describing the change of reference frame to the backwards direction, and the subscripts indicate the propagation direction. While it is possible to construct a gadget that implements the transpose by introducing time-reversal symmetry breaking elements~\cite{stromberg2024exploring}, here we instead exploit the fact that the transpose is a basis-dependent operation. More concretely, by adopting the convention $(S_1,S_2,S_3) \leftrightarrow (-Z,-Y,-X)$ for our Stokes parameters~\cite{frigo2022choice} we find that $P=\mathbb{1}$, and the polarization gadgets transform as the transpose under counterpropagation (see Methods). Superimposing two propagation directions through a gadget therefore allows us to implement the quantum time flip, with the photon path acting as a control degree of freedom. The specific coherent superposition of time flips in Eq.~\eqref{eq:time_flip2} is achieved through the use of fiber optic circulators.
    
The optical circuit in Fig.~\ref{fig:setup} begins with a bulk beam-splitter that initializes the control qubit into the state $\ket{+}_C = \frac{1}{\sqrt{2}}(\ket{0}_C+\ket{1}_C)$, after which two fiber circulators guide the photons through the $V$ gadget in two different directions, giving the joint control-target state
\begin{equation}
\frac{1}{\sqrt{2}}(
V\ket{\psi}_T \otimes \ket{0}_C + V^{T}\ket{\psi}_T \otimes \ket{1}_C).
\end{equation}

Entering the circulators from a different port, the photons are then directed to the $U$ gadget, which they once again propagate through in opposite directions, transforming the joint state to
\begin{equation}
\frac{1}{\sqrt{2}}(
 U^{T}V\ket{\psi}_T \otimes \ket{0}_C + UV^{T}\ket{\psi}_T \otimes \ket{1}_C).
\end{equation}
    
At the end of the optical circuit, a fiber beam-splitter applies a Hadamard gate on the control qubit, giving the state
\begin{equation} \small
    \bigg[\frac{UV^T+U^TV}{2}\bigg]\ket{\psi}_T |+\rangle_C  +  \bigg[\frac{UV^T-U^TV}{2}\bigg]\ket{\psi}_T  |-\rangle_C.
\end{equation} \normalsize 
A projective measurement on the control (path) qubit in the computational basis then reveals whether $(U,V)$ belong to $\mathcal{M}_+$ or $\mathcal{M}_-$. 
    
The partially common-path structure of the interferometer has two distinct advantages: (1) photons in the two different propagation directions of the interferometer hit exactly the same spots on the waveplates and the physical symmetries of the gadget therefore ensures the faithful implementation of the time flip independently of any imperfections in the waveplates, (2) the paths traversed in both directions do not contribute any phase noise to the interferometer, thereby simplifying the phase stabilization. More specifically, only the paths connecting the two beam-splitters with the fiber circulators, as well as the fibers directly between the circulators, add phase noise to the interferometer. These fiber components, as well as the bulk beam-splitter at the interferometer input, are housed in a thermally and acoustically insulated box. The passive stabilization of these elements is sufficient to bring the phase drift down to a value of approximately \SI{10}{\milli\radian\per\minute}. The use of a bulk-beamsplitter at the input was chosen in order to balance the losses induced by the fiber circulators through the free-space to fiber coupling, and to give control over the interferometer phase 
, through a piezo-electric actuator. This piezo was used to reset the phase of the interferometer prior to beginning the measurements, and was not employed for active feedback. The fiber beam-splitter at the output ensures perfect spatial mode overlap for high interferometric visibility.
    
\begin{figure}
        \includegraphics[width=1\linewidth]{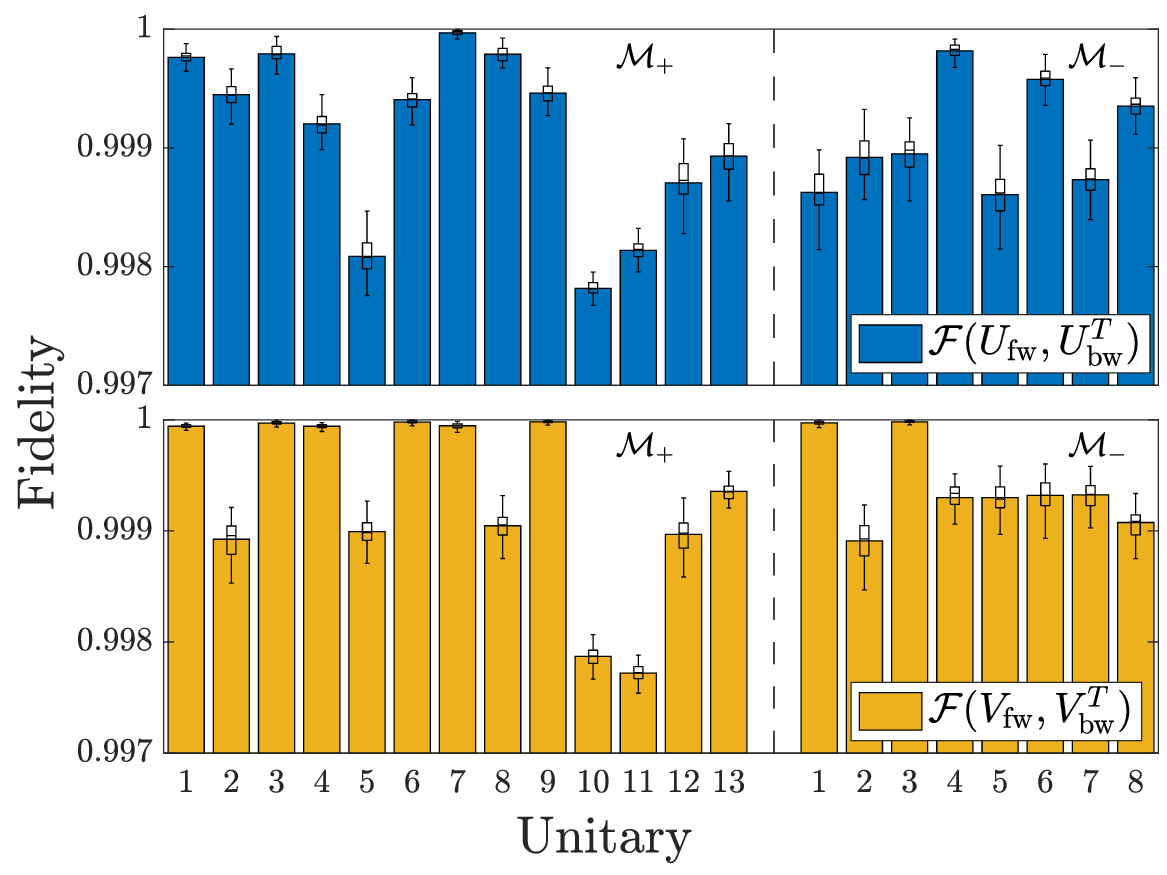}
        \centering
        \caption{\textbf{Unitary transposition fidelity.} The yellow and blue bars indicate the fidelity, $\mathcal{F}$, of the unitaries $U$ (top) and $V$ (bottom) from the sets $\mathcal{M}_+$ (left) and $\mathcal{M}_-$ (right), measured in the forward propagation direction, with respect to the transpose of the reconstructed unitary measured in the backwards propagation direction. Taller bars indicate a higher fidelity between the unitaries in the two propagation directions. The average fidelity is $0.9992 \pm 6.5\times 10^{-4}$ indicating that the gadgets faithfully implement the transpose. The uncertainties were estimated using a Monte-Carlo simulation of the tomography accounting for errors in the waveplate angles, and the superimposed box plot indicates the spread of the reconstructed fidelities. We attribute the residual errors to imperfect waveplate retardance in the tomography, and angle differences between the setting of the forwards and backwards unitaries, since in principle the gadgets perfectly implement the transpose of the unitary in the forward direction.}
            \label{fig:transposefid}
\end{figure}
   
\section{Results}
Before demonstrating the quantum time flip in the context of the game, we first verified the ability of a polarization gadget to implement both a unitary and its transpose simultaneously, in the two different propagation directions of the light. To this end, we performed quantum process tomography on the implemented unitaries from the sets $\mathcal{M}_+$ and $\mathcal{M}_-$, in both propagation directions. We then compared the fidelity $\mathcal{F} = \big\langle |(U_{\text{fw}}|\Psi\rangle)^{\dagger} U_{\text{bw}}^T | \Psi \rangle |^2\big\rangle_{\ket{\Psi}}$ between the reconstructed unitaries in the forward direction, $U_{\text{fw}}$ and $V_{\text{fw}}$, with the transposed reconstructed unitaries in the backwards direction, $U_{\text{bw}}^T$ and $V_{\text{bw}}^T$ (see Methods).
The results of this are shown in Fig.~\ref{fig:transposefid}. The average fidelity is greater than $0.999$, indicating that the gadgets correctly implement the transpose. Note that the fidelity of the transpose is independent of any errors in the retardance of the waveplates in the gadget itself. Such imperfections would cause the fidelity in the implementation of a desired unitary to drop, but would affect the forward and backward directions symmetrically. The same is true for undesired offsets in the waveplate angles, however in the measurements shown in Fig.~\ref{fig:transposefid} the unitaries in the two directions were measured in separate runs, causing them to indeed be sensitive to waveplate angle errors, in addition to errors in the tomography itself.
\begin{figure}[t]
\label{fig:box1}
        \begin{tcolorbox}[colback=teocolbox,left=1mm,top=1mm,boxrule=1pt,width=0.45\textwidth,halign=left]
            \fontsize{8}{8}
            \textbf{Box 1. Sets of unitary operators.}
            \vspace{-1pt}
            \begin{equation*}
            \begin{alignedat}{3}
                \mathcal{M}_+^{\mathrm{I}} =\{
                &(I, I), &&(I, X), &&(I, Z),\\ 
                &(X, I), &&(X, X), &&(X, Z),\\
                &(Z, I), &&(Z, X), &&(Z, Z)\}
            \end{alignedat}
            \quad
            \begin{aligned}
                \mathcal{M}_-^{\mathrm{I}} =\{
                &(Y, I), (Y, X), (Y, Z),\\
                &(I, Y), (X, Y), (Z, Y)\}\\
                &\vphantom{(I, Y)}
            \end{aligned}
            \end{equation*}
            \begin{equation*}
            \begin{aligned}
                \mathcal{M}_+^{\mathrm{II}} =\bigg\{
                &\bigg(\frac{X-Y}{\sqrt{2}}, \frac{X+Y}{\sqrt{2}}\bigg),
                \bigg(\frac{X+Y}{\sqrt{2}}, \frac{X-Y}{\sqrt{2}}\bigg),\\
                &\bigg(\frac{Z-Y}{\sqrt{2}}, \frac{Z+Y}{\sqrt{2}}\bigg),
                \bigg(\frac{Z+Y}{\sqrt{2}}, \frac{Z-Y}{\sqrt{2}}\bigg)
                \bigg\}
            \end{aligned}
            \end{equation*}
            \begin{equation*}
                \mathcal{M}_-^{\mathrm{II}} =\bigg\{
                \bigg(\frac{I+iY}{\sqrt2}, \frac{I-iY}{\sqrt{2}}\bigg), \bigg(\frac{I-iY}{\sqrt{2}}, \frac{I+iY}{\sqrt{2}}\bigg)
                \bigg\}
            \end{equation*}
            \vspace{5pt}
            \begin{equation*}
                \mathcal{M}_+ = \mathcal{M}_+^{\mathrm{I}}\cup \mathcal{M}_+^{\mathrm{II}}, \qquad
                \mathcal{M}_- = \mathcal{M}_-^{\mathrm{I}}\cup \mathcal{M}_-^{\mathrm{II}}
            \end{equation*}
        \end{tcolorbox}
\end{figure}
 \begin{figure*}
        \includegraphics[width=1\linewidth]{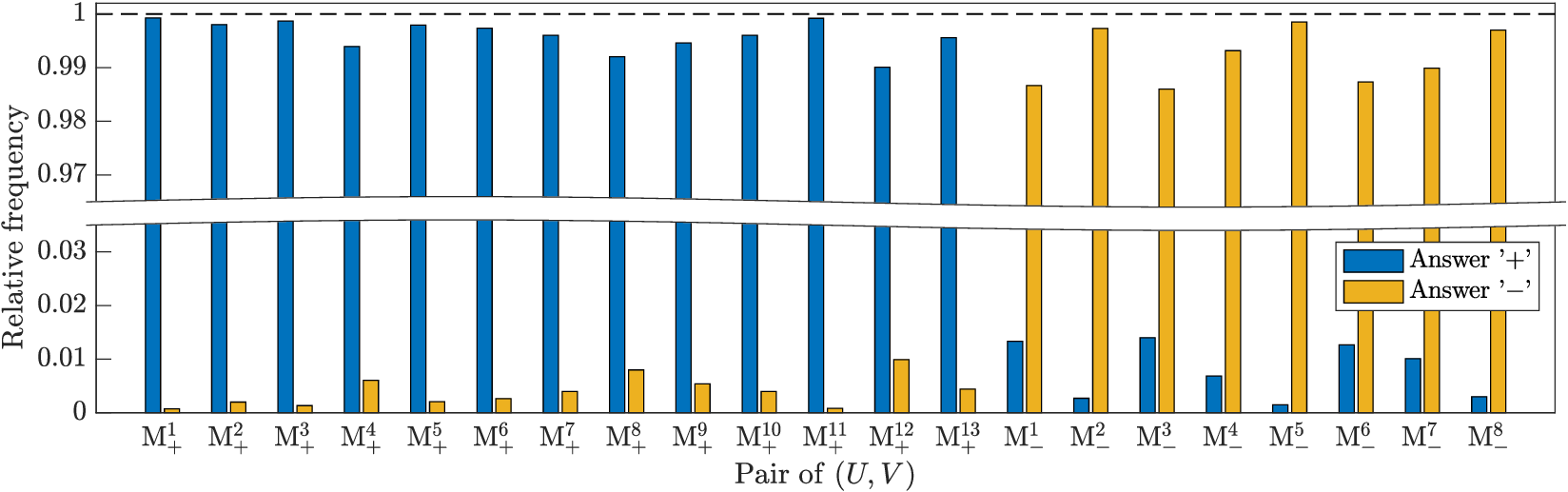}
        \centering
        \caption{\textbf{Observed relative outcome frequencies.} The figure shows the observed relative frequency of answers $f^{\mathrm{rel}}_{\pm}$ in the quantum flip game for all the pairs of unitaries in the sets $\mathcal{M}_+$ and $\mathcal{M}_-$. For the gates in the set $\mathcal{M}_+$ ($\mathcal{M}_-$) the game is won when the player outputs the answer `$+$' (`$-$'). The observed average winning frequency is $0.9945$. Since the bars corresponds to the actual number of times the different outcomes were recorded there is no associated uncertainty (see Methods).}
        \label{fig:barplot}
    \end{figure*}
Having verified the ability to implement a given unitary and its transpose with a single black box simultaneously, we then realised the game discussed in the previous sections. First, two-photon coincidence events for the different elements of $\mathcal{M}_+$ and $\mathcal{M}_-$ were collected sequentially to reduce the time spent rotating the waveplates. Second, the game itself was played using the collected data. In each round the referee uniformly randomly selects a pair of channels, and the player outputs an answer, `$+$' or `$-$', given by a unique two-photon event from the corresponding measurement set. Figure~\ref{fig:barplot} shows the relative frequencies $f^{\mathrm{rel}}_{\pm,k} = N^{\pm}_k/N_k$, where $N^{\pm}_k$ is the number of times the player output the answer `$\pm$' when the channels $(U_k,V_k)$ were picked, and $N_k$ is the total number of times these channels were selected by the referee. It can be seen that the player outputs the correct answer with a relative frequency higher than the indefinite tester bound of $0.92$ for every setting, and by extension any strategy that is separable in its time direction.
More specifically, the average winning frequency is found to be $0.9945$, with the best and worst case frequencies being $0.9993$ and $0.9860$, respectively.

The formulation of the indefinite-time-direction witness as a game with only two outcomes, win or lose, allows for a straightforward statistical interpretation of the results. Since we have an upper bound $p_\text{i.c.}\leq \frac{92}{100}$ on the probability of success for an indefinite tester , we can calculate the probability $P$ of such a player having obtained $v$ or more victories in $N$ rounds:
\begin{equation}
    P = \sum_{k=v}^N \binom{N}{k}p_{\mathrm{i.c.}}^k (1-p_{\mathrm{i.c.}})^{N-k}.
\end{equation}
This probability is exactly the $P$-value for the experimentally implemented process not being indefinite in its time direction. Out of the $N=10^6$ rounds played in the experiment, $v=994,512$ were won by successfully identifying the correct set, while $5,488$ rounds were lost.
Using a Chernoff bound tailored for the binomial distribution, we can provide an upper bound on the $P$-value, given by
\begin{equation}
\sum_{k=v}^N \binom{N}{k}p_{\mathrm{i.c.}}^k (1-p_{\mathrm{i.c.}})^{N-k}\leq \exp\left(-N\,D\left(\frac{v}{N}\big|\hspace{-.3mm}\big|p_{\mathrm{i.c.}}\right)\right),
\end{equation}
where $\exp$ is the exponential function and
\begin{equation}
D\left(\frac{v}{N}\big|\hspace{-.3mm}\big|p\right):=
\frac{v}{N}\ln\left(\frac{v}{Np}\right)+
\left(1-\frac{v}{N}\right)\ln\left(\frac{1-v/N}{1-p}\right)
\end{equation}
is the relative entropy. Direct calculation using $p_{\mathrm{i.c.}}=0.92$ shows that $D\left(\frac{v}{N}|\hspace{-.3mm}|p_{\mathrm{i.c.}}\right)\approx0.0627$, hence the $P$-value is upper bounded by $P\leq  e^{-10^4}$, which is an extremely small number. This rules out any explanation of the data in terms of convex mixtures of quantum processes that access the gates in a definite time direction. Since this is the defining characteristic for the class of processes with an indefinite time direction, we therefore conclude that the implemented process belongs to this class.

\section{Discussion}
In this work we have demonstrated, for the first time, a process that is inseparable in its time direction. Using an optical interferometer, we implemented a coherent superposition of arbitrary unitary transformations and their time-reversal. Such a process can only be probabilistically simulated by a quantum circuit with a definite time direction. Even agents equipped with two copies of the gates and able to combine them in an indefinite order cannot realise the process deterministically, unless they are given the ability of pre- and post-selecting quantum systems~\cite{oeckl2008general, oreshkov2016operational, chiribella2013quantum, silva2017connecting, svetlichny2011time, lloyd2011closed}.
It is worth noting that our implementation of controlled unitary transposition is not in contradiction with the no-go theorem, stating that there is no quantum circuit that can transform an unknown quantum unitary gate to its transpose~\cite{chiribella2020quantum, chiribella2021symmetries, quintino19PRA}. Our implementation adopts a device that implements a single-qubit gate $U$, and while this gate can remain unknown, the physical device itself is neither arbitrary nor unknown. Indeed, it is the particular symmetries of the physical device that necessarily and deterministically generate the transposed gate $U^T$.

While time itself does not flow backwards in any part of the experimental apparatus, our demonstration highlights the limitations of the quantum circuit model for describing the full range of quantum information processing protocols. This is analogous to the impossibility of perfect unitary coherent control within the quantum circuit model \cite{chiribella2020quantum, araujo2014quantum, thompson2018quantum, nakayama2014universal}.
Through a channel discrimination game, in which we outperform any strategy with a definite time direction, we furthermore certify that the coherent superposition of time directions yields a process that is inseparable in its time direction.

The study of indefinite causality led to the discovery and realisation of quantum information protocols with practical advantages~\cite{guerin2016exponential,wei2019experimental}, as well as a lively debate about the interpretation of these realisations~\cite{oreshkov2019time,paunkovic2020causal,fellous2022comparing,vilasini2022embedding}. We envision that future studies of processes with an indefinite time direction will similarly expand both the theoretical and experimental toolkit and open up new avenues for quantum information processing.
Indeed, a recent work has shown that processes with an indefinite time direction can show enhanced performance in certain communication tasks~\cite{liu2023quantum}, and the experimental methods presented here could be used to demonstrate such advantages.
We note that an experimental demonstration of an indefinite time direction was also presented in a parallel and independent work~\cite{guo2024experimental}.

In the context of future work we note that universal transposition of single-qubit gates is a sufficient building block for the transposition of multi-qubit gates, for instance using a Reck decomposition~\cite{reck1994experimental}, or through the inclusion of a reciprocal symmetric two-qubit gate~\cite{alonso2019quantum}.
We believe that the demonstration of coherent transposition of a two-qubit unitary using the former approach on a hyper-encoded two-qubit photonic state would be within experimental reach.
Finally, the investigation of time reversed quantum processes also holds applications in quantum thermodynamics. Indeed, in~\cite{chiribella2020quantum}, it was shown that the processes for which the quantum time flip produces another valid process are exactly those which can be written as linear combinations of unitary channels. That is, channels which do not decrease the entropy in either time direction. Nevertheless, the application of superpositions of two time directions in the context of thermodynamic work was recently studied in~\cite{rubino2021quantum,rubino2022inferring}.
Such superpositions could be realised using the quantum time flip by having it act on the unitary dynamics of the joint state-environment system.

\section{References}
%

\section*{Acknowledgements}
    We thank R. Peterson for fruitful discussions. This research was funded in whole, or in part, by the Austrian Science Fund (FWF) through [10.55776/F71] (BeyondC), [10.55776/FG5] (Research Group 5), and [10.55776/TAI483].
    M.T.Q. and C.B. acknowledge the Austrian Science Fund (FWF) through the SFB project BeyondC [10.55776/F71], a grant from the Foundational Questions Institute (FQXi) as part of the Quantum Information Structure of Spacetime (QISS) Project (qiss.fr). L.A.R. acknowledges support from the Erwin Schrödinger Center for Quantum Science \& Technology (ESQ Discovery). This project has received funding from the European Union’s Horizon 2020 research and innovation program under the Marie Skłodowska-Curie Grant Agreement No. 801110.
    The opinions expressed in this publication are those of the authors and do not necessarily reflect the views of the John Templeton Foundation. This project has received funding from the European Union’s Horizon 2020 research and innovation programme under the Marie Skłodowska-Curie grant agreement No 801110.
    It reflects only the authors' view, the EU Agency is not responsible for any use that may be made of the information it contains. ESQ has received funding from the Austrian Federal Ministry of Education, Science and Research (BMBWF).
\section*{Author contributions}
T.S. and P.S. carried out the experiment and collected the data. T.S. and I.A. build the experimental setup. T.S., P.S. and M.T.Q. analysed the data. T.S. and P.S. designed the experiment. M.T.Q. carried out the computer assisted proofs. M.A. and L.R. conceived the experiment. P.W., C.B. and I.A. supervised the project. All authors contributed to writing the manuscript.
\section*{Data availability}
All data used in this work is openly available in Zenodo under: \href{https://doi.org/10.5281/zenodo.7352614}{10.5281/zenodo.7352614}
\section*{Code availability}
The code used to perform the computer assisted proofs is openly available at the following online repository: \href{https://github.com/mtcq/UnitaryTransposition}{https://github.com/mtcq/UnitaryTransposition}

\onecolumngrid
\vspace{1.5\baselineskip}
\PRLsep
\section*{Methods}
    \subsection{Arbitrary unitary transposition}
        The description of linear retarders depends on the convention used for the polarization states, i.e. which Pauli matrices are associated with which Stokes parameters. The most commonly used convention in quantum optics is:
        \begin{equation}
            \label{eq:convention1}
            (S_1,S_2,S_3) \leftrightarrow (Z,X,Y),
        \end{equation}
        corresponding to the $\{H,V\}$, $\{+,-\}$ and $\{L,R\}$ polarizations being the eigenstates of $Z$, $X$ and $Y$ respectively. Under this convention, a linear retarder, such as a waveplate, at an angle $\theta$ to the vertical axis, is described by the following matrix:
        \begin{equation}
            \begin{gathered}
                U(\theta) = e^{-\frac{i}{2}\theta Y}
                e^{-\frac{i}{2}rZ}
                e^{\frac{i}{2}\theta Y} \\=
                \begin{bmatrix}
                    \cos(\theta) & -\sin(\theta)\\
                    \sin(\theta) & \cos(\theta)
                \end{bmatrix}
                \begin{bmatrix}
                    e^{i\frac{r}{2}} & 0 \\
                    0 & e^{-i\frac{r}{2}}
                \end{bmatrix}
                \begin{bmatrix}
                    \cos(\theta) & \sin(\theta)\\
                    -\sin(\theta) & \cos(\theta)
                \end{bmatrix}
            \end{gathered}
        \end{equation}
        where $r$ is the retardance of the element. Note that the matrix $U(\theta)$ is symmetric since:
        \begin{equation}
            \begin{aligned}
                U(\theta)^T &= (e^{\frac{i}{2}\theta Y})^T
                (e^{-\frac{i}{2}rZ})^T
                (e^{-\frac{i}{2}\theta Y})^T \\
                &= 
                e^{-\frac{i}{2}\theta Y}
                e^{-\frac{i}{2}rZ}
                e^{\frac{i}{2}\theta Y}.
            \end{aligned}
        \end{equation}
Propagating through such an element backwards has the effect of taking $\theta \mapsto -\theta$. This transformation can be written as:
\begin{equation}
    Z U(\theta) Z = U(-\theta)
\end{equation}
since 
\begin{equation}
    Z e^{-\frac{i}{2}\theta Y}Z = e^{\frac{i}{2}\theta Y}.
\end{equation}
For a general polarization gadget consisting of several linear retarders described by the unitary $U_{G,\mathrm{fw}}$ in the forwards direction we find the unitary for the backwards propagation direction, $U_{G,\mathrm{bw}}$, by transposing the order of the individual linear retarders and changing the sign of their respective angles:
        \begin{equation}
            U_{G,\mathrm{fw}} = U_1(\theta_1) \ldots U_n(\theta_n) \mapsto
            U_{G,\mathrm{bw}} = U_n(-\theta_n) \ldots U_1(-\theta_1)
        \end{equation}
        which can be written:
        \begin{equation}
            \label{eq:ztranspose}
            U_{G,\mathrm{fw}} \mapsto Z U_{G,\mathrm{fw}}^T Z
        \end{equation}
        since:
        \begin{equation}
            \begin{aligned}
                Z (U_1(\theta_1) \ldots U_n(\theta_n))^T Z
                &= 
                Z U_n(\theta_n) \ldots U_1(\theta_1) Z \\
                &= 
                Z U_n(\theta_n) Z \ldots Z U_1(\theta_1) Z\\
                &= U_n(-\theta_n) \ldots U_1(-\theta_1).
            \end{aligned}
        \end{equation}
        The transformation in Eq.~\eqref{eq:ztranspose} is not useful for realising the transpose, since the $Z$ gates around the unitary $U_{G,\mathrm{fw}}^T$ have to be undone to recover the transpose.
        
        However, this problem can be overcome by picking a different convention for the polarization basis states, such as $(S_1,S_2,S_3) \leftrightarrow (X,Y,Z)$ which is a cyclic permutation of the aforementioned one (corresponding to a rotation of the basis vectors by $\pi/3$ around the vector $\begin{bmatrix}
        1 & 1 & 1\end{bmatrix}$), and which is commonly used in polarimetry. In this work, we chose the convention:
        \begin{equation}
            \label{eq:convention2}
            (S_1,S_2,S_3) \leftrightarrow (-Z,-Y,-X).
        \end{equation}
        The minus signs are necessary to preserve the handedness of the coordinate system when exchanging $X$ and $Y$. That this convention yields the desired transformation under counterpropagation can be realised by noting that the Stokes parameters of a unitary always transform as $(S_1,S_2,S_3) \mapsto (S_1,-S_2,S_3)$, however for completeness we will perform the calculation explicitly. In the convention of Eq.~\eqref{eq:convention2} a linear retarder at an angle $\theta$ is written as:
        \begin{equation}
            U(\theta) = e^{\frac{i}{2}\theta X}
            e^{\frac{i}{2}rZ}
            e^{-\frac{i}{2}\theta X}
        \end{equation}
        and the corresponding unitary in the backwards direction is
        \begin{equation}
            \begin{aligned}
                U(-\theta) &= e^{-\frac{i}{2}\theta X}
                e^{\frac{i}{2}rZ}
                e^{\frac{i}{2}\theta X} \\
                &=
                (e^{\frac{i}{2}\theta X}
                e^{\frac{i}{2}rZ}
                e^{-\frac{i}{2}\theta X})^T \\
                &= U(\theta)^T.
            \end{aligned}
        \end{equation}
        It then follows that a general waveplate gadget also transforms as the transpose:
        \begin{equation}
            \begin{aligned}
                U_{G,\mathrm{fw}} = U_1(\theta_1) \ldots U_n(\theta_n) \mapsto
                U_{G,\mathrm{bw}} &= U_n(-\theta_n) \ldots U_1(-\theta_1) \\
                &=
                (U_1(\theta_1) \ldots U_n(\theta_n))^T\\
                &= U_{G,\mathrm{fw}}^T.
            \end{aligned}
        \end{equation}

One could alternatively get around the problem with Eq.~\eqref{eq:ztranspose} by introducing two more polarization gadgets implementing $Z$ operators on either side of the gadget in Eq.~\eqref{eq:ztranspose}, and making sure that these additional gadgets only act on one propagation direction. For example, by physically displacing the beam paths of the two propagation directions, so that the gadgets act on different spatial modes in the different propagation directions. This would, however, change the interpretation of the experiment with respect to the implementation in the main text, since the transformations in the two propagation directions would no longer be related by a physical symmetry. Instead they would depend on the transformations realised by the additional gadgets.

\subsection{Obtaining upper bounds for different classes of strategies}
We now detail how to obtain an upper bound on the winning probability of the game described in the main manuscript.
Let $N$ be the total number of pairs of unitary operators contained in the set $\mathcal{M}_+$ and  $\mathcal{M}_-$. Following a uniform distribution, i.e., with probably $1/N$, the referee picks a pair of unitary operators $(U_i,V_i)$. The player should then employ a quantum strategy to guess whether $(U_i,V_i)$ belongs to $\mathcal{M}_+$ or $\mathcal{M}_-$. Let $p(\pm|(U_i,V_i))$ the probability that the player guesses $(U_i,V_i)\in\mathcal{M}_\pm$. The probability of such player to win the game is then given by 
\begin{equation}
    p= \frac{1}{N}\Biggl(\sum_{(U_i,V_i)\in\mathcal{M}_+} p\big(+|(U_i,V_i)\big) + \sum_{(U_i,V_i)\in\mathcal{M}_-} p\big(-|(U_i,V_i)\big)\Biggr).
\end{equation}

For the qubit scenario considered here, we can analyse the case where unitary gates act backwards by simply considering the case where all involved unitary operators are transposed. This is true because, as discussed earlier, there are only two anti-homomorphisms from $\mathcal{SU}(d)$ to $\mathcal{SU}(d)$, and for any $U\in\mathcal{SU}(2)$, we have that $U^{-1}=\sigma_Y U^T \sigma_Y$. More explicitly, the winning probability for players using the unitary gates backwards is given by
\begin{equation}
    p= \frac{1}{N}\Biggl(\sum_{(U_i,V_i)\in\mathcal{M}_+} p\big(+|(U_i^T,V_i^T)\big) + \sum_{(U_i^T,V_i^T)\in\mathcal{M}_-} p\big(-|(U_i^T,V_i^T)\big)\Biggr).
\end{equation}
Also, as we show more explicitly later, since the success probability is linear function of the strategies, convex combinations of forward and backwards strategies cannot increase the maximal success probability. Hence it is enough to analyse the forward and backwards case.
    
When the player is restricted to  parallel strategies, the most general approach consists of preparing a quantum state $\rho$, sending part of this state to the operators $U_i$ and $V_i$, and then performing a quantum measurement with outcomes labelled as $+$ or $-$, that is,
\begin{equation} \label{eq:PAR} \small
    p_\text{par}\big(\pm|(U_i,V_i)\big)=\tr{\Big[M_\pm \, \Big(U_i\otimes V_i \otimes \id)\rho(U_i^\dagger\otimes V_i^\dagger \otimes \id) \Big)\Big]},
\end{equation}  \normalsize
\normalsize
where $M_+,M_-\geq0$ are the POVM operators associated to the outcomes $+$ and $-$, see Fig.~2 in the main text for a pictorial illustration.

Parallel strategies may be analysed in the (parallel) tester formalism~\cite{chiribella2009theoretical,bavaresco21}, also known as process POVM~\cite{Ziman08}. Let us label the linear spaces corresponding to the input and output spaces as $\H_I$ and $\H_O$ respectively. We can then write $U_i\otimes V_i:\H_I\to\H_O$ with $\H_I\cong\H_O\cong\mathbb{C}_2\otimes\mathbb{C}_2$.
In the tester formalism, operations are viewed as states and Eq.~\eqref{eq:PAR} may be written as the generalized Born's rule. More formally, we have that:
\begin{equation} \label{eq:PAR2} 
    p_\text{par}\big(\pm| (U_i,V_i)\big)=\tr{\Bigl[T_\pm \, \dketbra{U_i\otimes V_i}{ U_i\otimes V_i}\Bigr]},
\end{equation} 
where\footnote{Here $\mathcal{L}(\H_I\otimes\H_O)$ denotes the set of linear operators from $\H_I\otimes\H_O$ (linear endomorphisms). }
$T_+,T_-\in \mathcal{L}(\H_I\otimes\H_O)$ are tester elements and 
${\dket{U_i\otimes V_i}\in}(\H_I\otimes\H_O)$ is the Choi vector of $U_i\otimes V_i$ defined as:
\begin{equation}
    \dket{U_i\otimes V_i}:=\sum_l \ket{l}\otimes \Big(U_i\otimes V_i \ket{l}\Big),
\end{equation}
where $\{\ket{l}\}$ is the computational basis for $\H_I$.
The operators $T_+$ and $T_-$ are parallel testers when $T_+, T_-\geq0$ and their sum respects:
\begin{equation}
    T_+ + T_- = \sigma_I \otimes \id_O,
\end{equation}
where $\sigma\in\mathcal{L}(\H_I)$ is a quantum state.
As shown in Refs.~\cite{chiribella2009theoretical,bavaresco21,Ziman08}, all parallel strategies as in Eq.~\eqref{eq:PAR} can be represented by testers such as those in Eq.~\eqref{eq:PAR2}, and \textit{vice versa}. Hence, when optimizing over all possible strategies, instead of considering all possible states $\rho$ and measurements $M_\pm$ as in Eq.~\eqref{eq:PAR}, we may optimize over all valid testers $T_\pm$ as in Eq.~\eqref{eq:PAR2}. 

One advantage of using the tester formalism, is that the maximal probability of winning the discrimination game can be written in terms of a semidefinite program (SDP) via the following optimisation problem:

\begin{align}
\label{eq:PAR_primal}
\max\; &  \frac{1}{N}\Bigg[
\sum_{(U_i,V_i)\in\mathcal{M}_+}\tr\Big(T_+\;\dketbra{U_i \otimes V_i}{ U_i\otimes V_i}\;\Big) + \sum_{(U_i,V_i)\in \mathcal{M}_-}\tr\Big(T_- \;\dketbra{U_i\otimes V_i}{ U_i\otimes V_i}\Big) 
\Bigg] \\
\text{s.t.: }& T_+,T_- \geq0 \\
             &T_+ + T_- = \sigma_I \otimes \id_O \\
             &\tr(\sigma)=1.
\end{align} \normalsize
Following the steps of Ref.~\cite{bavaresco21}, the dual problem is given by:

\vspace{-\baselineskip}
\begin{align}   \label{eq:PAR_dual}
\min\; &   \tr(C)/d_I \\
\text{s.t.: }
&\frac{1}{N}\sum_{(U_i,V_i)\in\mathcal{M}_+} \; \dketbra{U_i\otimes V_i}{ U_i\otimes V_i} \leq    C
\label{eq:par_dual1} \\ 
&\frac{1}{N}\sum_{(U_i,V_i)\in \mathcal{M}_-} \; \dketbra{U_i\otimes V_i}{ U_i\otimes V_i} \leq    C
\label{eq:par_dual2}\\
&\tr_O(C)= \tr_{IO}(C)\,\frac{\id_I}{d_I}, \label{eq:par_dual3}
\end{align} \normalsize
where $d_I$ is the dimension of $\H_I$ (for our particular problem, $d_I=4$).
By the definition of dual problem, if we find a linear operator $C$ satisfying the feasibility constraints of inequality~\eqref{eq:par_dual1}, inequality~\eqref{eq:par_dual2}, and Eq.~\eqref{eq:par_dual3}, the quantity $\tr(C)/d_I$ is an upper bound on the maximal success probability. In order to obtain a computer-assisted-proof upper bound with fraction of integers, we use standard and efficient floating-point arithmetic algorithms to solve the SDP, obtain an operator $C$ which satisfies the constraints of the dual problem and truncate it in such a way that the feasibility constraints are still satisfied. We refer to our online repository (see Code Availability) for an implementation of this procedure and to Ref.~\cite{bavaresco21} for a detailed explanation on how to perform the truncation step.

When the player is restricted to  causal strategies (also referred to as sequential strategies), the most general approach consists of preparing a quantum state $\rho$, sending part of this state to the operators $U_i$ (or to $V_i$), applying a quantum channel $\mathcal{E}$, then performing the operation $V_i$ (or $U_i$), and finally performing a quantum measurement with outcomes labelled as $+$ or $-$, that is:

\vspace{-\baselineskip}
\begin{align}
    p_\text{seq}(\pm&|(U_i,V_i))=\tr\Big[M_\pm \,(V_i \otimes \id) \mathcal{E}\Big(U_i\otimes \id\, \rho\, U_i^\dagger\otimes \id\Big)(V_i^\dagger \otimes \id)\Big]. \nonumber
\end{align}   \normalsize
Using the concept of sequential testers~\cite{chiribella2009theoretical,bavaresco21}, we can also write the problem of finding the optimal causal strategy as an SDP. Since there is a notion of causal order, we label the input and output space of the first operation as $\H_{I_1}$ and $\H_{O_1}$ respectively. Analogously, we use $\H_{I_2}$ and $\H_{O_2}$ for the second operations. If the player uses the operation $U_i$ first and $V_i$ second, we have that $U_i:\H_{I_1}\to\H_{O_1}$ and $V_i:\H_{I_2}\to\H_{O_2}$. Following Ref.~\cite{bavaresco21}, the primal and dual problem for causal strategies are respectively given by

\begin{align} \label{eq:SEQ_primal}
\max\; &  \frac{1}{N}\Bigg[
\sum_{(U_i,V_i)\in\mathcal{M}_+}\tr\Big(T_+\;\dketbra{U_i \otimes V_i}{ U_i\otimes V_i}\;\Big)+ \sum_{(U_i,V_i)\in \mathcal{M}_-}\tr\Big(T_- \;\dketbra{U_i\otimes V_i}{ U_i\otimes V_i}\Big) 
\Bigg] \\
\text{s.t.: }& T_+,T_- \geq0 \\
             &T_+ + T_- = W_{I_1O_1I_2} \otimes \id_{O_2} \\
             &\tr_{I_2}(W_{I_1O_1I_2}) = \sigma_{I_1} \otimes \id_{O_1} \\
             &\tr(\sigma)=1.
\end{align} \normalsize
and

\vspace{-\baselineskip}
\begin{align}   \label{eq:SEQ_dual}
\min\; &   \tr(C)/d_I \\
\text{s.t.: }
&\frac{1}{N}\sum_{(U_i,V_i)\in\mathcal{M}_+} \; \dketbra{U_i\otimes V_i}{ U_i\otimes V_i} \leq    C \\ 
&\frac{1}{N}\sum_{(U_i,V_i)\in \mathcal{M}_-} \; \dketbra{U_i\otimes V_i}{ U_i\otimes V_i} \leq    C\\
& \tr_{O_2}(C)= \tr_{I_2O_2}(C)\otimes \frac{\id_{I_2}}{d_{I_2}} \\ 
& \tr_{O_1I_2O_2}(C)= \tr_{I_1O_1I_2O_2}(C)\, \frac{\id_{I_1}}{d_{I_1}}. 
\end{align} \normalsize
Another sequential strategy would be to use $V_i$ before $U_i$. For this case, the semidefinite program is then exactly the same as the one before, but we exchange the roles of $V_i$ and $U_i$. Our methods show that, when $U_i$ precedes $V_i$, the success probability is bounded by $\frac{90}{100}\leq p_{UV}\leq \frac{91}{100}$, and when $V_i$ precedes $U_i$, the success probability is bounded by $\frac{90}{100}\leq p_{VU}\leq \frac{91}{100}$. Since the two bounds coincide, we have $\frac{90}{100}\leq p_\text{causal}\leq \frac{91}{100}$.

When the player is restricted to general quantum strategies without a definite causal order, the strategies are described by means of an indefinite tester~\cite{chiribella18cause_effect}, which consist of  positive semidefinite operators that add up to a process matrix~\cite{oreshkov2012quantum}, that is $T_+ + T_- =W$, where $W$ is a bipartite process matrix. Following Ref.~\cite{bavaresco21}, and defining the trace-and-replace maps as $_iX:=\tr_i(X)\otimes \id_i$, the primal and the dual problem are respectively given by

\vspace{-\baselineskip}
\begin{align} \label{eq:GEN_primal}
\max\; &  \frac{1}{N}\Bigg[
\sum_{(U_i,V_i)\in\mathcal{M}_+}\tr\Big(T_+\;\dketbra{U_i \otimes V_i}{ U_i\otimes V_i}\;\Big) \\ \nonumber
&\phantom{\frac{1}{N}\Bigg[ aaa a}+ \sum_{(U_i,V_i)\in \mathcal{M}_-}\tr\Big(T_- \;\dketbra{U_i\otimes V_i}{ U_i\otimes V_i}\Big) 
\Bigg] \\
\text{s.t.: }& T_+,T_- \geq0 \\
             &T_+ + T_- = W \\
             &_{I_2O_2}W = _{O_1I_2O_2}W \\
             &_{I_1O_1}W = _{O_2I_1O_1}W \\
             &W =_{O_1}W+_{O_2}W -_{O_1O_2}W \\
             &\tr(W)=\tr(\id_{O_1O_2}).
\end{align} \normalsize
and

\vspace{-\baselineskip}
\begin{align}   \label{eq:GEN_dual}
\min\; &   \tr(C)/d_I \\
\text{s.t.: }
&\frac{1}{N}\sum_{(U_i,V_i)\in\mathcal{M}_+} \; \dketbra{U_i\otimes V_i}{ U_i\otimes V_i} \leq    C \\ 
&\frac{1}{N}\sum_{(U_i,V_i)\in \mathcal{M}_-} \; \dketbra{U_i\otimes V_i}{ U_i\otimes V_i} \leq    C\\
&  _{O_1}C= _{I_1O_1}C \\ 
&  _{O_2}C= _{I_2O_2}C .
\end{align} \normalsize

\subsection{Data analysis}
As described in the main text, the game was played by having the referee pick pairs of unitaries from the sets $\mathcal{M}_{\pm}$ in a uniformly random way in every round. The player's outcome was determined by the first unused photon detection event in the event list corresponding to that choice of unitary by the referee. More concretely, let $O^{j,k}_{\pm}$ be the $k$-th element in the time-ordered list of detection events $\mathcal{O}^{j}_{\pm}$ for the implemented pair of unitaries $M_{\pm}^j$. Then the outcome of the $n$-th round of the game, in which the referee picked the pair of unitaries $M_{\pm}^j$ for the $k$-th time, is $O^{j,k}_{\pm}$.

During the course of this game the player outputs the answer `$+$' (`$-$') a total of $N_j^+$ ($N_j^-$) times in the $N_j$ rounds that the referee selects the pair of channels $(U_j,V_j)$. The relative frequencies with which the player outputs these answers can be written as:
\begin{equation}
    f^{\mathrm{rel}}_{+,j} = \frac{N^{+}_j}{N^j},\quad
    f^{\mathrm{rel}}_{-,j} = \frac{N^{-}_j}{N^j}.  
\end{equation}
These observed relative frequencies are shown in Fig.~5 in the main text. The values of these observed relative frequencies do not by themselves have an associated uncertainty, and are purely observed quantities. In many single photon experiments, quantities such as these are interpreted as empirical estimates of underlying probabilities, and such estimates do carry uncertainties. Even in perfect experimental realisations, finite counting statistics would introduce Poissonian noise in this type of estimation. However, the statistical method we use to determine the confidence in our conclusion - the calculation of the $P$-value - allows us to make statements about the underlying probability distribution without directly estimating it. Specifically, that its expectation value exceeds the bound imposed on the winning probability of any strategy with a definite time direction.

In order to filter out background events resulting from various back-reflections in the experimental setup, as well as detector dark counts, two-fold coincidence events between the signal and idler photons were used to time filter the detection events.

The superconducting nanowire detectors used in the experiment have a slight polarization dependence in their detection efficiency, and due to the different pairs of unitaries generating different target qubit states the event rates for different implemented unitaries varied. This difference in efficiency was not necessary to account for, because the number of events for each pair of unitaries was truncated, in reverse chronological order, to match the setting with the fewest events. To find the numbers of rounds won and lost, the data was sampled from once, drawing $10^6$ different samples from unique, chronologically ordered (for each setting) detection events. The exact number of won and lost rounds in this sampling were $994,512$ won and $5,488$ lost.

A detection efficiency imbalance is also present in the two output ports of the interferometer, corresponding to the two different measurement outcomes of the control qubit. This efficiency difference could quite easily be characterised and corrected for, however such actions are equivalent to classical post-processing and is captured by the indefinite tester. Imbalanced detection efficiency could therefore not lead to a violation of the bound, and is not necessary to correct for since the data already violates the bound. This is a different way of stating the semi-device independence of our methods.

The measurement of the fidelity between the unitary implemented in one direction and the transpose of the unitary in the other direction was performed with coherent light. To estimate the fidelity, the two unitaries were first fitted to the data using a maximum likelihood estimation and then the fidelity was calculated by evaluating the following average:
\begin{equation}
    \mathcal{F} = \big\langle |(U_{\text{fw}}|\Psi\rangle)^{\dagger} U_{\text{bw}}^T | \Psi \rangle |^2\big\rangle_{\ket{\Psi}},
\end{equation}
taken over 1000 Haar-random states $\ket{\Psi}$. This was done in every step of a Monte-Carlo simulation to estimate the measurement uncertainties induced by the waveplate errors.

\subsection{Semi-device independence of demonstration}
In this section we will elaborate on what is meant by our certification methods being semi-device independent. Our usage of this term is consistent with the notion of semi-device independence introduced in~\cite{bavaresco19}. That our demonstration is semi-device independent means that the measurement that the player performs does not have to be characterised. Equivalently, the player does not have to trust that their measurement device implements a specific measurement. It is a statement about the required assumptions on the measurement.

The basis for the claim that our demonstration is semi-device independent lies in the fact that the derivation of the bounds for the strategies depicted in Fig.~2.a-c in the main text included an optimization over all possible binary measurements the player could perform. This means that there is no measurement that a player using these strategies could perform that would allow them to violate the bounds we derived. Hence, a violation of these bounds has the same interpretation regardless of what measurements the player performed.

It is worth noting that semi-device independence does not imply that the ability of the player to violate the bounds is independent of the measurement they perform. Indeed, measurement imperfections can reduce the winning rate of the player. This can cause them to fail to certify that they employ a certain strategy, even if they do in fact employ that strategy.

A concrete consequence of the semi-device independence is that imperfections in the measurement do not need to be accounted for, and the measurement itself does not need to be modelled in the data analysis. This is in contrast to device-dependent methods, which rely on well characterised measurements to draw conclusions about the observed results. A device-dependent verification method that frequently appears in experimental quantum information science is the witness operator, for example entanglement witnesses or causal witnesses. Such witness operators can also be constructed for the task described in the main text. A witness operator $\hat{S}$ can be used to certify a certain statement about a quantum system or process by experimentally evaluating its expectation value, and confirming that it satisfies some bound:
\begin{equation}
\label{eq:witness}
    \langle \hat{S} \rangle < B.
\end{equation}
Empirically evaluating $\langle \hat{S} \rangle$ requires the witness operator to be decomposed in terms of experimentally measurable observables, and the expectation values of these observables to be estimated. Imperfections in the measurement devices induce uncertainties in these estimates, which in turn propagate as uncertainties into the expectation value of the witness operator. A statistically significant violation of the inequality \eqref{eq:witness} therefore requires well characterised measurement devices.

\end{document}